\documentclass[twocolumn,nofootinbib,amsmath,amssymb,a4paper]{revtex4}

\usepackage{graphicx}
\usepackage{dcolumn}
\usepackage{bm}

\begin{document}

\title{QCD factorisation and flavour symmetries illustrated in
$B_{d,s}\to KK$ decays}

\author{S. Descotes-Genon}
 \email{Sebastien.Descotes-Genon@th.u-psud.fr}
\affiliation{
Laboratoire de Physique
Th\'eorique, 
CNRS/Univ. Paris-Sud 11 (UMR 8627),
91405 Orsay Cedex, France}

\begin{abstract}
We present a new analysis of $B_{d,s}\to KK$ modes within the SM, 
relating them in a controlled way through $SU(3)$-flavour symmetry 
and QCD-improved factorisation. We propose a set of sum rules for 
$B_{d,s}\to K^0{\bar K}^0$ observables. We determine $B_s\to KK$
branching ratios and CP-asymmetries as functions of 
$A_{dir}(B_d \to K^0{\bar K}^0)$, with a good agreement with current
experimental measurements of CDF. Finally, we predict the amount of 
$U$-spin breaking between $B_d\to \pi^+\pi^-$ and $B_s\to K^+K^-$. 
\end{abstract}

\maketitle

The current data in $B$-physics suggests that $B_d$-decays agree well with
SM predictions, while $B_s$-decays remain poorly known and
might be affected by New Physics. Within the Standard Model, the CKM
mechanism correlates the electroweak part of these transitions, but 
quantitative predictions are difficult due to hadronic effects. The
latter can be estimated relying on the approximate $SU(3)$-flavour 
symmetry of QCD :
information on hadronic effects, extracted from data in one channel, 
can be exploited in other channels related by flavour symmetry, 
leading to more accurate predictions within the Standard Model.

In addition to isospin symmetry, an interesting theoretical tool
is provided by $U$-spin symmetry, which relates $d$- and $s$-quarks. 
Indeed, this symmetry holds for long- and short-distances
and does not suffer from electroweak corrections, 
making it a valuable instrument to analyse processes with significant 
penguins and thus a potential sensitivity to New Physics. 
However, due to the significant difference $m_s-m_d$, 
$U$-spin breaking corrections of order 30~\% may occur,
depending on the processes.

As a first application of $U$-spin, relations were obtained between
$B_d\to \pi^+\pi^-$ and 
$B_s\to K^+ K^-$. This led to correlations among the observables
in the two decays such as branching ratios and CP 
asymmetries~\cite{Fleischer:1999pa,Fleischer:2002zv} and to
a prediction for 
$Br(B_s\to K^+ K^-)=(35^{+73}_{-20})\cdot 10^{-6}$~\cite{Buras:2004ub}.
These results helped to investigate the potential
of such decays to discover New Physics~\cite{London:2004ej,Baek:2005wx}.
Unfortunately, the accuracy of the method is limited not only by the
persistent discrepancy between Babar and Belle on
$B_d\to\pi^+\pi^-$ CP asymmetries, but also 
by poorly known $U$-spin corrections. In these analyses,
the ratio of tree contributions
$R_c=|T^s_{K\pm}/T^d_{\pi\pm}|$ was taken from 
QCD sum rules as $1.76\pm 0.17$, updated
to $1.52^{+0.18}_{-0.14}$~\cite{Khodjamirian:2003xk}. In addition,
the ratio of penguin-to-tree ratios
$\xi=|(P^s_{K\pm}/T^s_{K\pm})/(P^d_{\pi\pm}/T^d_{\pi\pm})|$
was assumed equal to $1$~\cite{Buras:2004ub} 
or $1\pm 0.2$~\cite{London:2004ej,Baek:2005wx}
in agreement with rough estimates within QCD 
factorisation (QCDF)~\cite{Safir:2004ua}. Recent updates
on $U$-spin methods were given during this workshop~\cite{talks}.

QCDF may complement flavour symmetries by
a more accurate study of short-distance effects. However, this
expansion in $\alpha_s$ and $1/m_b$ cannot predict 
some significant $1/m_b$-suppressed long-distance effects, which
have to be estimated through models.
Recently, it was proposed to combine QCDF and $U$-spin in 
the decays mediated by penguin operators $B_d\to K^0 \bar{K}^0$ and 
$B_s\to K^0 \bar{K}^0$~\cite{DGMV}.

The SM amplitude for a $B$ decaying into two mesons can be split into
tree and penguin contributions~\cite{B}:
\begin{equation}
\bar{A}\equiv A(\bar{B}_q\to M \bar{M})
  =\lambda_u^{(q)} T_M^{qC} + \lambda_c^{(q)} P_M^{qC}\,,
\end{equation}
with $C$ denoting the charge of the decay products, and the
products of CKM factors $\lambda_p^{(q)}=V_{pb}V^*_{pq}$. Using
QCDF~\cite{BBNS,BN}, one can perform a $1/m_b$-expansion of the
amplitude.
The tree and penguin contributions in $\bar{B}_s\to K^+K^-$ and
$\bar{B}_s\to K^0 \bar K^0$ in QCDF are :
\begin{eqnarray}
{\hat T^{s\,\pm}} &=&
  \bar\alpha_1 + \bar\beta_1\\ \nonumber
&& \  + \bar\alpha^u_4 + \bar\alpha^u_{4EW} + \bar\beta_3^u + 2\bar\beta_4^u
      - \frac{1}{2} \bar\beta^u_{3EW} + \frac{1}{2} \bar\beta^u_{4EW}
  \\
{\hat P^{s\,\pm}} &=&
      \bar\alpha^c_4 + \bar\alpha^c_{4EW} + \bar\beta_3^c + 2\bar\beta_4^c
      - \frac{1}{2} \bar\beta^c_{3EW} + \frac{1}{2} \bar\beta^c_{4EW}\\
\label{eq3}
{\hat T^{s\, 0}} &=&
  \bar\alpha_4^u-\frac{1}{2}\bar\alpha_{4EW}^u
    +\bar\beta_3^u +2 \bar\beta_4^u - \frac{1}{2} \bar\beta^u_{3EW} - \bar\beta^u_{4EW}
\\  \label{eq4}
{\hat P^{s\, 0}} &=&
 \bar\alpha_4^c-\frac{1}{2}\bar\alpha_{4EW}^c
    +\bar\beta_3^c +2 \bar\beta_4^c
    - \frac{1}{2} \bar\beta^c_{3EW} - \bar\beta^c_{4EW}
\end{eqnarray}
where $\hat P^{sC}=P^{sC}/A^s_{KK}$, $\hat T^{sC}=T^{sC}/A^s_{KK}$
and $A^q_{KK}=M^2_{B_q} F_0^{\bar{B}_q\to K}(0) f_K
{G_F}/{\sqrt{2}}$. The superscripts identify the channel and the
bar denotes quantities for decays with a spectator $s$-quark. The
tree and penguin contributions ${T^{d\, 0}}$ and $P^{d\, 0}$ for
$\bar{B}_d\to K^0 \bar K^0$ have the same structure as
eqs.~(\ref{eq3}) and (\ref{eq4}), with unbarred $\alpha$'s and
$\beta$'s recalling the different nature of the spectator
$d$-quark.

At NLO in $\alpha_s$, $\alpha$'s are linear combinations of vertex
corrections, hard-spectator 
 terms and penguin
contractions, whereas $\beta$'s are sums of annihilation
contributions. The weights of the various contributions
are expressed in terms of $\alpha_s$ and Wilson
coefficients~\cite{BN}. $\alpha$'s and
$\beta$'s contain the two most significant terms in the $1/m_b$
expansion: the LO terms, dominated by short distances, and the NLO
terms in $1/m_b$ that include the potentially large long-distance
corrections. The latter, parameterised in QCDF through quantities
denoted $X_H$ (in power corrections to the hard-scattering part of
$\alpha_i$) and $X_A$ (in the annihilation parameters $\beta_i$),
are singled out since they may upset the quick convergence of the
$1/m_b$ expansion. The other $1/m_b$-suppressed contributions,
dominated by short distances, are under control and small, i.e,
leading to a ${\cal O}(5-10\%)$ error.
In~\cite{DGMV}, we showed that comparing $B_d$- and
$B_s$-decays into the same final states helps to cancel the potentially
large long-distance
$1/m_b$-suppressed effects ($X_{A,H}$), yielding improved
SM predictions. 

\begin{figure}
\includegraphics[width=8cm]{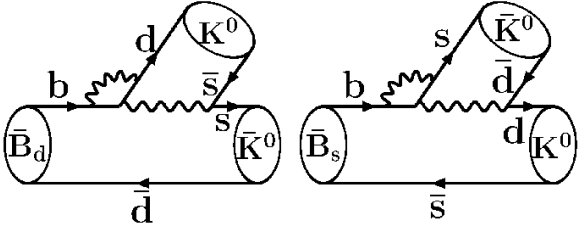}
\caption{Diagrams contributing to $\bar{B}_d\to K^0\bar{K}^0$
(left) and $\bar{B}_s\to K^0\bar{K}^0$ (right) related through
$U$-spin transformations. \label{fig:uspink0}}
\end{figure}

\section{Sum rules} 

Let us start with the difference 
$\Delta_d\equiv T^{d0}-P^{d0}$  is free from the troublesome
NLO infrared-divergence (modelled by $X_{A,H}$) that may be enhanced numerically by
the chiral
factor $r_\chi^K=2m_K^2/m_b/m_s$ from twist-3 distribution amplitudes.
Hard-scattering ($X_H$) and annihilation ($X_A$) terms
occur in both penguin and tree contributions, but remarkably they
cancel in the short-distance difference:
\begin{eqnarray}
\Delta_d&=&A_{KK}^d[\alpha_4^{u}-\alpha_4^{c} +
\beta_3^{u}-\beta_3^{c}+2\beta_4^{u}-2\beta_4^{c}]\\\nonumber
&=&
A_{KK} \times \alpha_s C_F C_1 \times [\bar{G}(m_c^2/m_b^2)-\bar{G}(0)]/(4\pi N_c)
\end{eqnarray}
neglecting (small) electroweak contributions. The function
$\bar{G}=G_K-r_\chi^K \hat{G}_K$ combines one-loop
integrals from the penguin terms $P_4$ and $P_6$ defined in
Sec 2.4 in ref.~\cite{BN}. The same cancellation
of long-distance $1/m_b$-corrections happens for
$\Delta_s \equiv T^{s0}-P^{s0}$. Taking into account
the uncertainties coming from the QCDF inputs \cite{BN},
we get
$
\Delta_d=(1.09\pm 0.43) \cdot 10^{-7} + i (-3.02 \pm 0.97) \cdot
10^{-7} {\rm GeV}$ and
$
\Delta_s=(1.03 \pm 0.41) \cdot 10^{-7} + i (-2.85\pm 0.93) \cdot
10^{-7}  {\rm GeV}.
$

These two theoretical quantities can be related to observables,
namely the corresponding branching ratio and coefficients of the
time-dependent CP-asymmetry:
\begin{eqnarray}
&&\frac{\Gamma(B_d(t)\to K^0\bar{K}^0)-\Gamma(\bar{B}_d(t)\to K^0\bar{K}^0)}
     {\Gamma(B_d(t)\to K^0\bar{K}^0)+\Gamma(\bar{B}_d(t)\to K^0\bar{K}^0)}\\
\nonumber
&&\qquad \qquad      =
  \frac{A_{dir}^{d0} \cos(\Delta M\cdot t) + A_{mix}^{d0} \sin(\Delta
M\cdot t)}
       {\cosh(\Delta\Gamma_d t/2)- A_{\Delta}^{d0}\sinh(\Delta\Gamma_d
t/2)}\,,
\end{eqnarray}
where we define~\cite{Fleischer:1999pa}:
$A_{dir}^{d0}= ({|A|^2-|\bar{A}|^2})/({|A|^2+|\bar{A}|^2})$,
$A_{\Delta}^{d0}+i A_{mix}^{d0}=
-({2 e^{-i \phi_d} A^* \bar{A}})/({|A|^2+|\bar{A}|^2})$ and
$\phi_d$ the phase of $B_d-\bar{B}_d$ mixing.
$A_{\Delta}^{d0}$ is unlikely to be measured due to the small
width difference $\Delta\Gamma_d$, but it can be obtained from the
other asymmetries by means of the relation
$|A_{\Delta}^{d0}|^2+|A_{dir}^{d0}|^2+|A_{mix}^{d0}|^2=1$.

One can derive the following relation for $B_d\to K^0 \bar{K}^0$:
\begin{eqnarray} \label{eq:srd}
|\Delta_d|^2&\!=\!&{\frac{BR^{d0}}{L_d}}
\{x_1 + [x_2 \sin\phi_d - x_3 \cos\phi_d] A_{mix}^{d0}\\
&&\qquad\qquad\qquad
   -[x_2 \cos\phi_d + x_3 \sin\phi_d ] A_{\Delta}^{d0} \}\,, \nonumber
\end{eqnarray}
where $L_d=\tau_d \sqrt{M_{Bd}^2 - 4 M_K^2}/(32 \pi M_{Bd}^2)$ and:
\begin{eqnarray*}
x_1&=&[|\lambda_c^{(d)}|^2 + |\lambda_u^{(d)}|^2 -
          2 |\lambda_c^{(d)}| |\lambda_u^{(d)}| \cos \gamma]/n^2\,,\\
x_2&=& -[|\lambda_c^{(d)}|^2 + |\lambda_u^{(d)}|^2 \cos 2\gamma-
              2 |\lambda_c^{(d)}| |\lambda_u^{(d)}| \cos\gamma]/n^2 \,,\\
x_3&=& -[1 - \cos\gamma \times |\lambda_u^{(d)}|/|\lambda_c^{(d)}|]/n\,,
\end{eqnarray*}
with $n={2 |\lambda_c^{(d)}| |\lambda_u^{(d)}| \sin \gamma}$.
A similar relation between $\Delta_s$ and $B_s\to K^0\bar{K}^0$
observables is obtained by replacing
$|\lambda_u^{(d)}| \to |\lambda_u^{(s)}|$, $|\lambda_c^{(d)}| \to
-|\lambda_c^{(s)}|$, and $d\to s$ for all indices.

These sum rules can be used either 
as a way to extract the SM value of one observable (say $|A_{dir}^{s0}|$)
in terms of the two others ($BR^{s0}$ and $A_{mix}^{s0}$) and $\Delta_s$,
as a SM consistency test
between $BR^{s0}$, $|A_{dir}^{s0}|$ and $A_{mix}^{s0}$
(and similarly for the $B_d^0 \to K^0 {\bar K}^0$ observables), or
as  a way of determining CKM parameters~\cite{dlim}.
These relations are free from the long-distance power-suppressed
model-dependent quantities
$X_A$ and $X_H$  that are a main error source in the direct computation of
$A_{dir}^{s0}$ within QCDF.


\begin{figure}
\includegraphics[width=4cm]{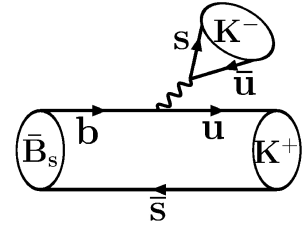}
\caption{Diagram contributing to 
$\bar{B}_s\to K^+\bar{K}^-$ from a tree operator,
without counterpart in $\bar{B}_d\to K^0\bar{K}^0$.}
\label{fig:tree}
\end{figure}

\section{Flavour symmetries and QCDF}

Using $U$-spin symmetry, we can relate the two penguin-mediated decays $\bar{B}_d \to K_0\bar{K}_0$ and
$\bar{B}_s \to K_0\bar{K}_0$, as exemplified in fig.~\ref{fig:uspink0}
(see also ref.~\cite{flrec} in relation to $B\to \pi\pi$). 
$U$-spin breaking should be much smaller
here than usual: it does not
affect final-state interaction since both decays involve the same outgoing
state, and it shows up mainly in
power-suppressed effects. This is confirmed by QCDF:
\begin{eqnarray} \nonumber
P^{s0}&=&f P^{d0}
  \Big[1+(A^d_{KK}/P^{d0})\Big\{\delta\alpha_4^c
           -\delta\alpha_{4EW}^c/2\\ \nonumber
&&\quad
  +\delta\beta_3^c+2\delta\beta_4^c
  -\delta\beta_{3EW}^c/2
  -\delta\beta_{4EW}^c
    \Big\}\Big]\,,\\
T^{s0}&=&f T^{d0}
  \Big[1+(A^d_{KK}/T^{d0})\Big\{
  \delta\alpha_4^u-\delta\alpha_{4EW}^u/2\\ \nonumber
  &&\quad
   +\delta\beta_3^u
   +2\delta\beta_4^u
   -\delta\beta_{3EW}^u/2
   -\delta\beta_{4EW}^u
    \Big\}\Big]\,,
\end{eqnarray}
where we define the $U$-spin breaking differences
$\delta\alpha_i^p\equiv\bar\alpha_i^p-\alpha_i^p$ (id. for $\beta$).
Apart from the factorisable ratio~:
\begin{equation*}
f={A_{KK}^s}/{A_{KK}^d}
  ={M_{B_s}^2 F_0^{\bar{B}_s\to K}(0)}/{[M_{B_d}^2 F_0^{\bar{B}_d\to K}(0)]}
\end{equation*}
which should be computed on the lattice,
$U$-spin breaking arises through $1/m_b$-suppressed
contributions in which most long-distance contributions have cancelled out.

First,
the hard-spectator scattering ($\delta\alpha$) probes
the difference between $B_d$- and $B_s$-distribution amplitudes
which is expected small, since
the dynamics of the heavy-light meson in the limit $m_b\to\infty$ should vary
little from $B_d$ and $B_s$.
Second, the annihilation contributions ($\delta\beta$) contain a $U$-spin breaking
part when the gluon is emitted from the light quark in the $B_{d,s}$-meson
(this effect from $A_1^i$ and $A_2^i$ defined in~\cite{BN}
is neglected in the QCDF model for annihilation terms).
Taking the hadronic parameters in~\cite{BN}, we obtain
$|P^{s0}/(fP^{d0})-1| \leq 3 \%$ and $|T^{s0}/(fT^{d0})-1| \leq 3 \%.$

Relations exist between $\bar{B}_d \to K_0\bar{K}_0$ and
$\bar{B}_s \to K^+ K^-$ as well.
A combination of $U$-spin and isospin rotations
leads from the penguin contribution
in $\bar{B}_d \to K_0\bar{K}_0$ to
that in $\bar{B}_s \to K_0\bar{K}_0$,
then to $\bar{B}_s \to K^+ K^-$,
up to electroweak corrections (it corresponds to
fig.~\ref{fig:uspink0} up to replacing $d\to u$ in the right-hand diagram).
On the other hand, there are no such relations between trees,
since $\bar{B}_s \to K^+ K^-$ contains tree contributions (see
fig.~\ref{fig:tree}) which have no counterpart in the penguin-mediated decay
$\bar{B}_d \to K_0\bar{K}_0$. This is seen in QCDF as well:
\begin{eqnarray}
&&\!\!\!\!\!\!P^{s\pm}=f P^{d0}\Big[
    1 + \frac{A_{KK}^d}{P^{d0}}
      \Big\{\frac{3}{2}(\alpha^c_{4EW}+\beta_{4EW}^c)
       +\delta\alpha_4^c
\nonumber \\
&&\!\!\!
  +\delta\alpha_{4EW}^c
  +\delta\beta_3^c+2\delta\beta_4^c
  -\frac{1}{2}(\delta\beta_{3EW}^c-\delta\beta_{4EW}^c)
  \Big\}\Big]  \,, \\
&&\!\!\!\!\!\!\frac{T^{s\pm}}{A_{KK}^s\bar\alpha_1}=
 1 + \frac{T^{d0}}{A_{KK}^d \bar\alpha_1}
     + \frac{1}{\bar\alpha_1}\Big\{\bar\beta_{1}+
       \frac{3}{2}(\alpha^u_{4EW}+\beta^u_{4EW})
\nonumber \\
&&\!\!\!
     + \delta\alpha_4^u+\delta\alpha_{4EW}^u
     + \delta\beta_3^u+2\delta\beta_4^u
       -\frac{1}{2}(\delta\beta_{3EW}^u-\delta\beta_{4EW}^u) \Big\}\,.
\nonumber
\end{eqnarray}
Terms are ordered in decreasing size (in particular,
curly brackets in $T^{s\pm}$ should be tiny). From QCDF, we obtain
the following bounds: $|P^{s\pm}/(fP^{d0})-1| \leq 2 \% $ and
$|T^{s\pm}/(A^s_{KK} \bar\alpha_1)-1-T^{d0}/(A^d_{KK} \bar\alpha_1)| \leq  4 \%$.
The latter shows that flavour-symmetry breaking corrections are
smaller than $T^{d0}/(A^d_{KK} \bar\alpha_1)=O(10\%)$. Fortunately, $T^{s\pm}$
is strongly CKM suppressed in $B_s\to K^+ K^-$ so that the
uncertainty on its QCDF determination will affect
the branching ratio and CP-asymmetries only marginally.

\begin{table*}
\begin{center}
{\footnotesize
\begin{tabular}{|l||c|c|c||c|c|c|}
\hline
                       &  $BR^{s0}\,\times 10^6$       &
                       $A_{dir}^{s0}\,\times 10^2$  &  $A_{mix}^{s0}\,\times
                       10^2$
                       &  $BR^{s\pm}\,\times 10^6$       &
                       $A_{dir}^{s\pm}\,\times 10^2$  &  $A_{mix}^{s\pm}\,\times 10^2$
    \\
\hline
$A_{dir}^{d0}=-0.2$    & $ 18.4\pm 6.5 \pm 3.6$  & $ 0.8 \pm 0.3 $  & $ -0.3  \pm 0.8$
                       & $ 21.9\pm 7.9 \pm 4.3$  & $ 24.3\pm 18.4$  & $ 24.7 \pm 15.5$  \\
\hline
$A_{dir}^{d0}=-0.1$    & $ 18.2\pm 6.4 \pm 3.6$  & $ 0.4 \pm 0.3 $  & $ -0.7 \pm 0.7$
                       & $ 19.6\pm 7.3 \pm 4.2$  & $ 35.7\pm 14.4 $ & $ 7.7 \pm 15.7$       \\
\hline
$A_{dir}^{d0}=0$       & $ 18.1\pm 6.3 \pm 3.6$  & $ 0 \pm 0.3 $    & $ -0.8 \pm 0.7$
                       & $ 17.8\pm 6.0 \pm 3.7$  & $ 37.0\pm 12.3$  & $ -9.3\pm 10.6$         \\
\hline
$A_{dir}^{d0}=0.1$     & $ 18.2\pm 6.4 \pm 3.6$  & $ -0.4 \pm 0.3$  & $ -0.7 \pm 0.7$
                       & $ 16.4\pm 5.7 \pm 3.3$  & $ 29.7\pm 19.9$  & $ -26.3 \pm 15.6 $       \\
\hline
$A_{dir}^{d0}=0.2$     & $ 18.4\pm 6.5 \pm 3.6 $ & $ -0.8 \pm 0.3$  & $ -0.3 \pm 0.8$
                       & $ 15.4\pm 5.6 \pm 3.1$  & $ 6.8 \pm 28.9$  & $ -40.2 \pm 14.6 $            \\
\hline
\end{tabular}
}
\end{center}
\caption{Observables for $\bar{B}_s\to K^0\bar{K}^0$ and
$\bar{B}_s\to K^+K^-$ as functions of the direct asymmetry
$A_{dir}(\bar{B}_d\to K^0\bar{K}^0)$ within the SM. We take
$\lambda_u^{(d)}=0.0038 \cdot e^{-i\gamma}$,
$\lambda_c^{(d)}=-0.0094$, $\lambda_u^{(s)}=0.00088\cdot
e^{-i\gamma}$, $\lambda_c^{(s)}=0.04$,
 and $\gamma=62^\circ$,
$\phi_d=47^\circ$, $\phi_s=-2^\circ$~\cite{CKMfitter}.
 \label{TableSMResults}
}
\end{table*}

\section{SM predictions for $B_s\to KK$ decays}

The dynamics of $B_d\to K^0\bar{K}^0$ involves
three hadronic real parameters
(modulus of the tree, modulus of the penguin and relative
phase) which we can pin down through three observables:
$BR^{d0}$, $A_{dir}^{d0}$ and $A_{mix}^{d0}$.
Only $BR^{d0}=(0.96\pm 0.25)\cdot 10^{-6}$~\cite{brd} has been measured.
However the direct asymmetry
$A_{dir}^{d0}$ should be observable fairly easily (for instance,
$A_{dir}^{d0}=0.19\pm 0.06$ in QCDF) whereas the mixed asymmetry is likely
 small ($A_{mix}^{d0}=0.05\pm 0.05$ in QCDF).

If only $A_{dir}^{d0}$ becomes available,
we have only 2 experimental constraints for 3 hadronic parameters.
Then we may exploit a theoretically well-controlled QCDF constraint
to get $T^{d0}$ and $P^{d0}$ from $BR^{d0}$,
$A_{dir}^{d0}$ and the QCDF value of $\Delta_d\equiv T^{d0}-P^{d0}$,
free from infrared divergences and thus with little model dependence.
This system yields two constraints in the complex pla\-ne ($x_P,y_P$) for
$P^{d0}$ : a circular ring and a diagonal strip~\cite{DGMV}, which can be
satisfied only if $|A_{dir}^{d0}|<0.2$, and then yield two different 
solutions with opposite signs for
${\rm Im\ } P^{d0}$, yielding two solutions for $(P^{d0},T^{d0})$.

From the measured
value of the branching ratio for $B_d\to K^0\bar{K}^0$, and choosing a particular value of the
direct asymmetry $A_{dir}^{d0}$, we get the penguin and tree
contributions as explained above. Then, the bounds in II yield the hadronic
parameters in $B_s\to KK$ decays up to small uncertainties. To be more conservative,
we actually stretch the bounds in II relating
$B_d$ and $B_s$ hadronic parameters up to 5~\% in order to account for
well-behaved short-distance $1/m_b$-corrections not yet included.

We obtain observables
as functions of $A_{dir}^{d0}$ in Table~\ref{TableSMResults}.
In the case
of the branching ratios, we have split the error in two parts. The first
uncertainty
comes from the QCDF estimates of $\Delta_d$ and $\bar\alpha_1$, the
theoretical constraints derived in II to
relate $B_d$ and $B_s$ decays and the measurement of $BR^{d0}$
(this experimental uncertainty dominates the others).
The second error stems from (factorisable) $U$-spin breaking
terms: $f=0.94 \pm 0.2$ (cf.~\cite{BN}).

Table \ref{TableSMResults} corresponds only to the solution of the constraints
 with ${\rm Im\ } P^{d0}>0$.
But $BR^{d0}$, $A_{dir}^{d0}$ and $\Delta_d$ yield
two different solutions for $(T^{d0},P^{d0})$, and thus
for $(T^{s\pm},P^{s\pm})$. Only one solution is physical, whereas
the other stems from the non-linear nature of the constraints. We can
lift this ambiguity by exploiting
a channel related to $B_s\to K^+K^-$ through $U$-spin, namely
$\bar{B}_d\to \pi^+\pi^-$~\cite{Fleischer:1999pa,Buras:2004ub,LM,London:2004ej,Baek:2005wx}. As
explained in ref.~\cite{DGMV}, this allows us to reject one
of the two solutions, and to compute for the other
the $U$-spin breaking parameters :
\begin{equation}
R_C=\left|\frac{T^{s\pm}}{T^{d\pm}_{\pi\pi}}\right|
  =2.0\pm 0.6 \qquad
\xi=\left|\frac{P^{s\pm}}{T^{s\pm}}\frac{P^{d\pm}_{\pi\pi}}{T^{d\pm}_{\pi\pi}}\right|=0.8\pm
  0.3
\end{equation}

The determination of $BR^{s\pm}$ is improved
compared to the $U$-spin
extraction from $\bar{B}_d\to \pi^+\pi^-$ \cite{Buras:2004ub,London:2004ej,Baek:2005wx}, and in
good agreement with the recent CDF measurement~\cite{cdf}: 
\begin{eqnarray}
\left. BR^{s\pm}\right|_{exp} \cdot 10^6  &=&  24.4 \pm 1.4 \pm 4.6 
\end{eqnarray}

\section{Conclusions} 

We have combined experimental data, flavour
symmetries and QCDF to propose sum rules for
$B_{d,s} \to K^0 \bar{K}^0$ observables and to
give SM constraints on $B_s\to K\bar{K}$ in a controlled way.
Tree ($T^{d0}$) and penguin ($P^{d0}$) contributions to 
$B_d\to K^0 \bar{K}^0$ can be determined by 
combining the currently available data with
$|T^{d0}-P^{d0}|$, which can be accurately computed in QCDF because
long-distance effects, seen as infrared divergences, cancel in this difference.
$U$-spin suggests accurate relations between these hadronic parameters in
$B_d\to K^0 \bar{K}^0$ and those in $B_s\to K^0 \bar{K}^0$. Actually,
we expect similar long-distance effects since the $K^0 \bar{K}^0$ final state
is invariant under the $d$-$s$ exchange. Short distances
are also related since the two processes are mediated by penguin operators
through diagrams with the same topologies. 
$U$-spin breaking arises only in a few places : factorisable corrections 
encoded in
$f=[M_{B_s}^2 F^{B_s\to K}(0)]/[M_{B_d}^2 F^{B_d\to K}(0)]$, 
and non-factorisable corrections from weak annihilation and 
spectator scattering. 

Because of these expected tight relations, QCDF can be relied upon
to assess $U$-spin breaking between the two decays. Indeed, up to
the factorisable factor $f$, penguin (as well as tree) contributions to
both decays are numerically very close. Penguins in 
$B_d\to K^0 \bar{K}^0$ and $B_s\to K^+ K^-$ should have very close values as 
well, whereas no such relation exists for the (CKM-suppressed) tree 
contribution to the latter.

 These relations among hadronic parameters,
inspired by $U$-spin considerations and 
quantified within QCD factorisation, can be exploited to predict :
$Br(B_s\to K^0 \bar{K}^0) = (18\pm 7\pm 4\pm 2)\cdot 10^{-6}$ and
$Br(B_s\to K^+ \bar{K}^-) = (20\pm 8\pm 4\pm 2)\cdot 10^{-6}$, in
very good agreement with the latest CDF measurement. 
The same method provides significantly improved
determinations of the $U$-spin breaking ratios $\xi=0.8\pm 0.3$ and
$R_c=2.0\pm 0.6$. These results have been exploited to determine
the impact of supersymmetric models on these decays~\cite{Baek:2006pb}.

Our method merges ingredients from flavour symmetries and QCD factorisation
in order to improve the accuracy of the predictions. Flavour symmetries
must rely on global fudge factors typically of order 30\%, without providing
hints where symmetry breaking is large or small (typically, $\xi$ is
guesstimated, and not computed). QCD factorisation allows one
to classify the contributions of the various operators of the effective
Hamiltonian, but it suffers from a model-dependence in potentially large
$1/m_b$ corrections. We use QCD factorisation to determine the
short-distance contributions, and we replace models for long distances 
by experimental pieces of information from other decays related by 
flavour symmetries. Both methods are exploited optimally to yield
accurate predictions for $B_s\to KK$ as functions of $B_d$ observables.

If sizeable NP effects occur, the SM correlations between $B_d$
and $B_s$ decays exploited here should be broken, leading to
departure from our predictions. 
New results on $B\to K$ form factors and
on the $B_d\to K^0 \bar{K}^0$ branching ratio and direct CP-asymmetry 
should lead to a significant 
improvement of the SM predictions in the $B_s$ sector.
The potential of other pairs of nonleptonic $B_d$ and $B_s$ decays remains
to be investigated.

\begin{acknowledgments}
I thank J. Matias and J. Virto for a very enjoyable collaboration. 
This work was supported in part by the 
EU Contract No. MRTN-CT-2006-035482, \lq\lq FLAVIAnet''.
\end{acknowledgments}

\end{document}